\documentclass[a4paper,11pt]{article}
\pdfoutput=1 

\usepackage{jheppub} 

\usepackage[T1]{fontenc} 

\usepackage{graphicx} 

\usepackage{booktabs} 
\usepackage{array} 
\usepackage{paralist} 
\usepackage{verbatim} 
\usepackage{subfig} 

\usepackage{fancyhdr} 
\usepackage{sectsty}
\allsectionsfont{\sffamily\mdseries\upshape} 

\usepackage[nottoc,notlof,notlot]{tocbibind} 
\usepackage[titles,subfigure]{tocloft} 

\usepackage{amsfonts}

\usepackage{amsmath}
\usepackage{amssymb}

\title{\boldmath Five-loop anomalous dimensions of $\phi^Q$ operators in a scalar theory with $O(N)$ symmetry}

\author[a]{Qingjun Jin,}
\author[a]{Yi Li}

\affiliation[a]{Graduate School of China Academy of Engineering Physics, No. 10 Xibeiwang East Road, Haidian District, Beijing, 100193, China\\ }

\emailAdd{qjin@gscaep.ac.cn}
\emailAdd{yili@gscaep.ac.cn}

\abstract{We compute the complete $Q$-dependence of anomalous dimensions of traceless symmetric tensor operator $\phi^Q$ in $O(N)$ scalar theory to five-loop. The renormalization factors are extracted from  $\phi^Q\rightarrow Q\phi$ form factors, and the integrand of form factors are constructed with the help of unitarity cut method.  The anomalous dimensions match the known results in \cite{Badel:2019oxl, Antipin:2020abu}, where the leading and subleading terms in the large $Q$ limit were obtained using a semiclassical method. 

}

\begin{document} 
\maketitle
\flushbottom

\section{Introduction}
Different physical systems around second order phase transitions exhibit universal critical behaviors, and can be studied through field-theoretic renormalization group approach \cite{Wilson:1971bg, Wilson:1971dh}.
The critical exponents can be obtained from the scaling dimensions of operators at the Wilson-Fisher fixed point \cite{Wilson:1971dc} combined with resummation methods \cite{LeGuillou:1979ixc}.
One of the most extensively studied quantum field theories in this context is the scalar $\phi^4$ theory with $O(N)$ symmetry, which describes critical phenomena of different statistical systems including  superfluid $\sideset{^4}{}{\mathop{He}}$, ferromagnets, etc.

Recently, a semiclassical method (see e.g. \cite{Son:1995wz, Hellerman:2015nra, Alvarez-Gaume:2016vff, Alvarez-Gaume:2019biu, Arias-Tamargo:2019xld, Gaume:2020bmp}) was developed in \cite{Badel:2019oxl}, which provides an convenient way to study the anomalous dimensions of $\phi^Q$-type charged operators in $\phi^4$ theory with $U(1)$ symmetry. The leading and subleading terms of the anomalous dimensions in the large $Q$ limit can be evaluated to arbitrarily high loops.
The method was applied to the $O(N)$ scalar theory in \cite{Antipin:2020abu}, and the result was consistent with the traditional perturbative calculation to 3-loop. The comparison was extended to 4-loop in \cite{Jack:2021ypd}, where the subset of Feynman diagrams which contribute to the anomalous dimensions in the large charge limit were evaluated. 
Similar computations and comparisons have also been performed for a $U(N) \times U(N)$ model in \cite{Antipin:2020rdw, Jack:2021lja}, and for a $U(N) \times U(M)$ model in \cite{Antipin:2021akb}.

In order to further test the anomalous dimensions from semiclassical method, the perturbative computations should be extended to five and more loops. 
In the perturbative approach, the anomalous dimensions are extracted from the ultraviolet (UV) divergences of correlation functions or form factors. 
However, there are two major challenges in this computation. 
First, the number of contributing Feynman diagrams increase drastically as the loop number and $Q$ increase.
Second, it can be very difficult to evaluate the UV divergences of 5-loop integrals in the form factors.

The first challenge can be overcome by using the unitarity cut method, which has been playing a key role in the computation of gauge and gravity scattering amplitudes and form factors (see e.g. \cite{Bern:1994zx, Bern:2008qj, Boels:2008ef,Bern:2012uc,Cachazo:2013hca,Cachazo:2013iea, Bern:2015ooa,Yang:2016ear}).
The form factor with arbitrary $Q$ can be constructed by a few simple "irreducible" cut integrands.
The UV divergences are computed using the UV decomposition method \cite{Jin:2022zcj}, which direct extracts the UV counterterms from the integrand.


In this work, we perform a perturbative computation of the anomalous dimensions of $\phi^Q$-type operators in the $O(N)$ model to 5-loop.
In Section~\ref{section:unitarity}, we construct the form factor integrands using unitarity cut method.
In Section~\ref{section:anomalous}, we present the full $Q$-dependence of anomalous dimensions to 5-loop, and compare 
with known results.
In Section~\ref{section:discussion}, we discuss some possible future directions.

\section{Construction of $\phi^Q\rightarrow Q\phi$ form factors using unitarity cut}
\label{section:unitarity}

The $\phi^4$ theory with $O(N)$ symmetry is given by the Lagrangian,
\begin{equation}
\mathcal{L}=\frac{1}{2}(\partial_{\mu}\phi)^2+\frac{g}{4!}(\phi^2)^2\ , 
\end{equation}
in which $\phi=(\phi_1,\cdots,\phi_N)$ is a real scalar field with $N$ components.
We will work in a $D=4-\epsilon$ dimensional Euclidean space\footnote{As far as $\phi^4$ theory is concerned, there is no essential difference between Euclidean space and Minkowski space, since the theory is free of IR divergence.}, and use the minimal subtraction scheme.
Following \cite{Antipin:2020abu}, we consider the symmetric traceless component of $\phi^Q$ operator:
\begin{equation}
\begin{aligned}
(\phi^2)_{i_1i_2}&\equiv \phi_{i_1}\phi_{i_2}-\frac{1}{N}\delta_{i_1i_2}\phi^2,\ \\
(\phi^3)_{i_1i_2i_3}&\equiv \phi_{i_1}\phi_{i_2}\phi_{i_3}
-\frac{1}{N+2}\phi^2(\phi_{i_1}\delta_{i_2i_3}+\phi_{i_2}\delta_{i_1i_3}
+\phi_{i_3}\delta_{i_1i_2}),\ \\ 
\cdots&\\
\end{aligned}
\end{equation}
The anomalous dimensions of $\phi^Q$ operators with $Q=2,4$ were obtained in \cite{Braun:2013tva, Kompaniets:2019zes} and \cite{Calabrese:2002bm} respectively, and in this paper we consider operators with general $Q$. The anomalous dimensions of these operators can be extracted from the UV divergences of the following form factor, 
\begin{equation}
\Bigl(\mathbf{F}_Q\Bigr)_{i_1\cdots i_Q}^{j_1\cdots j_Q}
\equiv \int d^Dxe^{-ik\cdot x}\Bigl\langle \phi_{j_1}(p_1)\cdots \phi_{j_Q}(p_Q)\Bigr|(\phi^Q)_{i_1\cdots i_Q}(x)\Bigr|0\Bigr\rangle\ .
\end{equation}
We will call it the $\phi^Q\rightarrow Q\phi$ form factor.

In Section \ref{subsection:O(N)}, we study the structure of $O(N)$ indices, and show that these indices can be removed by contracting the operator and external states with the same zero-norm vector. 
In Section \ref{subsection:unitarity}, we construct form factor integrands with the help of unitarity cut method. A few "irreducible" cut integrands are taken as building blocks of form factor integrands with general $Q$.
In Section \ref{subsection:2n-amplitude}, we discuss the scattering amplitudes which are required in the unitarity cut computations.

\subsection{$O(N)$ tensor structures}
\label{subsection:O(N)}

The first difficulty we encountered in the computation of these form factors comes from the cumbersome $O(N)$ tensor structure in the operator. In the simplest case $Q=2$, the Feynman rule already contains 3 terms,
\begin{equation}
V_{i_1i_2}^{j_1j_2}=\delta_{i_1}^{j_1}\delta_{i_2}^{j_2}
+\delta_{i_1}^{j_2}\delta_{i_2}^{j_1}
-\frac{2}{N}\delta_{i_1i_2}\delta^{j_1 j_2}\ .
\end{equation}
The number of terms in the Feynman rule increases exponentially as $Q$ increases.

In order to overcome this difficulty, we observe all Feynman diagrams contributing to the $\phi^Q\rightarrow Q\phi$ form factor actually have the same tensor structure, which equals the corresponding Feynman rule:
\begin{equation}
\Bigl(\mathbf{F}_Q\Bigr)_{i_1\cdots i_Q}^{j_1\cdots j_Q}
=V_{i_1\cdots i_Q}^{j_1\cdots j_Q}\mathcal{F}_Q(p_1,\cdots,p_n) \ ,
\end{equation}
in which $\mathcal{F}_Q(p_1,\cdots,p_n)$ is a loop integrand without $O(N)$ indices.
The factorization of the tensor structure is basically a consequence of charge conservation, but it  can also be understood on the level of Feynman diagrams. A proof of this factorization based on Feynman diagrams can be found in Appendix \ref{appendixA}.

This factorization allows us to contract the operator with a zero-norm complex vector $q$ satisfying $q^2=0$ and $q\cdot \bar{q}=1$, without altering the anomalous dimension\footnote{The most simple choice of $q$ is $\frac{1}{\sqrt{2}}(1,i,0,\cdots, 0)$. In the $U(1)$ case ($N=2$), $\mathcal{O}_Q\propto (\phi_1+i\phi_2)^Q$ is exactly the operator $\phi^n$ ($n$ corresponds to $Q$ here) in \cite{Badel:2019oxl}.}:
\begin{equation}
\mathcal{O}_Q
\equiv\frac{1}{Q!}q_{i_1}\cdots q_{i_Q}(\phi^Q)_{i_1\cdots i_Q}
=\frac{(q\cdot\phi)^Q}{Q!}\ .
\end{equation}
The Feynman rule of $\mathcal{O}_Q$ only contains a single term. The form factor of $\mathcal{O}_Q$ is
\begin{equation}
\int d^Dxe^{-ik\cdot x}\Bigl\langle \phi_{j_1}(p_1)\cdots \phi_{j_Q}(p_Q)\Bigr|\mathcal{O}_Q(x)\Bigr|0\Bigr\rangle
=\bar{q}_{j_1}\cdots \bar{q}_{j_Q}\mathcal{F}_Q(p_1,\cdots ,p_Q)\ ,
\end{equation}
in which $\bar{q}$ appears instead of $q$ because the operator is "in-coming" instead of "out-going". 

The expression can be further simplified if we also contract the external fields $\phi_{j}$ with $q$, and define $\varphi\equiv q\cdot \phi$. The form factor is now free of $O(N)$ indices,
\begin{equation}
\int d^Dxe^{-ik\cdot x}\Bigl\langle \varphi(p_1)\cdots \varphi(p_Q)\Bigr|\mathcal{O}_Q(x)\Bigr|0\Bigr\rangle
=\mathcal{F}_Q(p_1,\cdots,p_Q)\ .
\end{equation}
We will call it the $\varphi^Q\rightarrow Q \varphi$ form factor.
The $L$-loop contribution to the $\varphi^Q\rightarrow Q \varphi$ form factor will be denoted by $\mathcal{F}_Q^{(L)}$. The tree form factor is simply\footnote{Strictly speaking, all form factors contain a $\delta^D(k-\sum_{i=1}^Qp_i)$ from the $x$-integration, which guarantees the momentum conservation. So $\mathcal{F}_Q^{(0)}=1$ actually means $\mathcal{F}_Q^{(0)}=\delta^D(k-\sum_{i=1}^Qp_i)$.
For compactness we will neglect this overall factor in all tree and loop form factors. }
$\mathcal{F}_Q^{(0)}=1$.

\subsection{Unitarity cut}
\label{subsection:unitarity}
Unitarity based methods simplify the computation of scattering amplitudes(form factors) in two ways. First, the intermediate expressions during the computation can be substantially simplified by making use of on-shell amplitudes, instead of off-shell interaction vertices, as the building block of loop amplitude integrand. Second, different cut channels can be related by permutations of the internal and external legs, so a lot repeated computations can be avoided.

Since in this work we are interested in a pure scalar theory, the on-shell constraint does not really simplify the expressions, but unitarity cut method can still help us to exploit the permutation symmetry of the internal and external legs.

\begin{figure}[htb]
\centering
\includegraphics[scale=0.5]{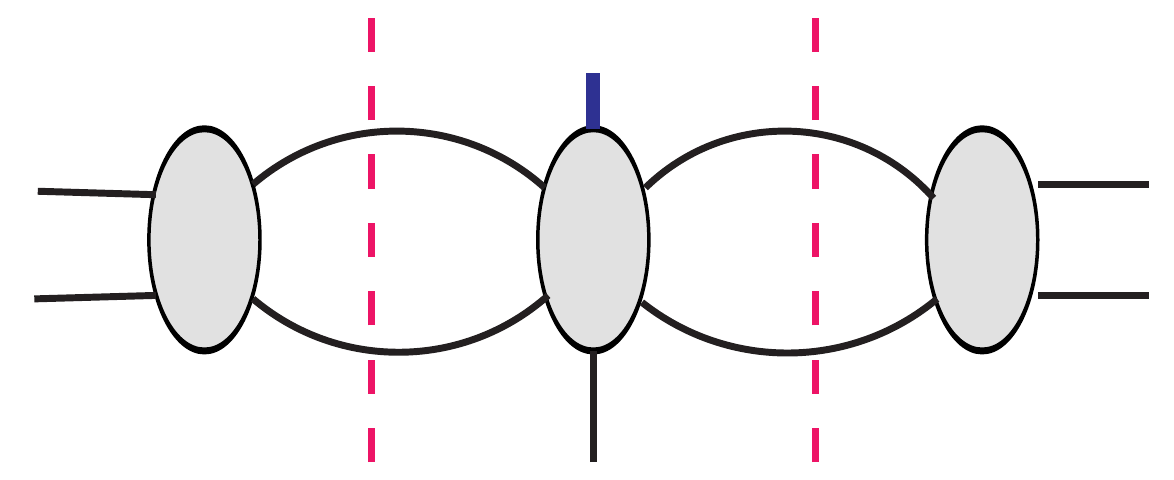}
\caption{A reducible cut. The blue line is the $\varphi^Q$ operator.  Each solid blob represents a tree amplitude (or tree form factor), while a blob with several holes represents a multiloop amplitude (or multiloop form factor).}
\label{fig:phi4cut2}
\end{figure}

\begin{figure}[htb]
\centering
\includegraphics[scale=0.5]{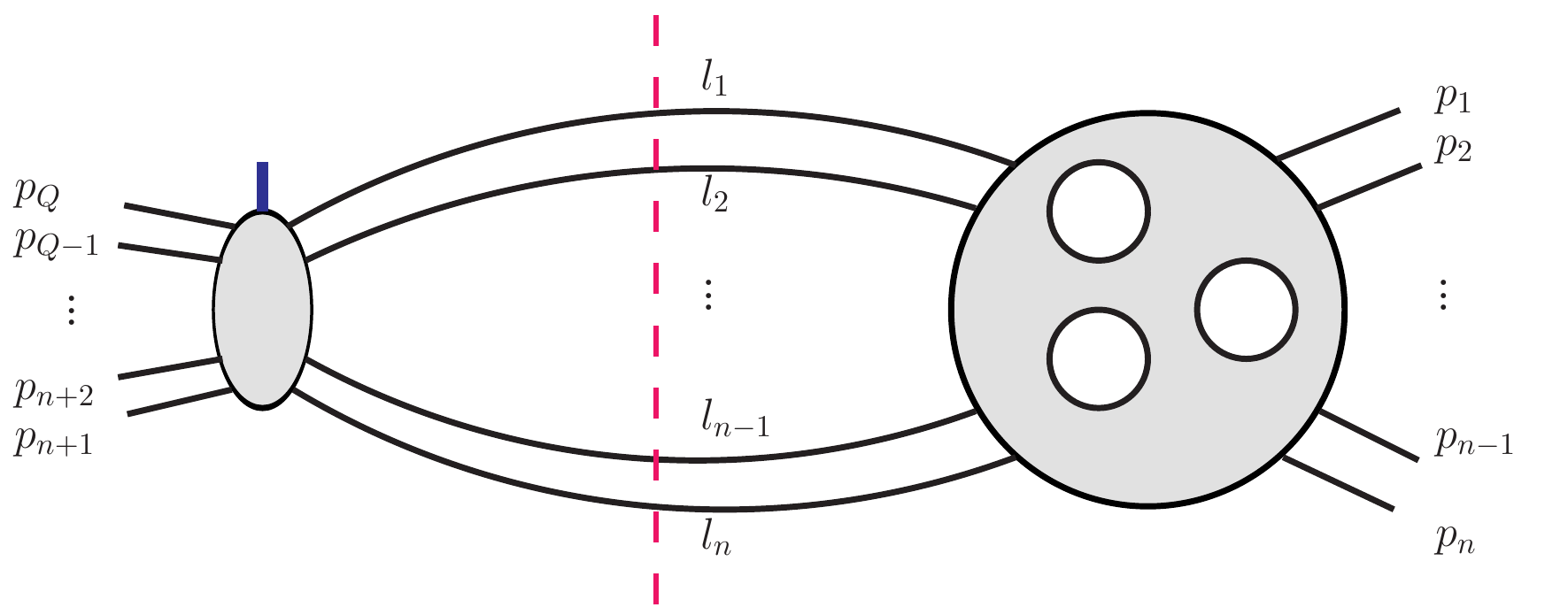}
\caption{An irreducible $n$-line cut of $\varphi^Q\rightarrow Q \varphi$ $L$-loop form factor. The l.h.s. of the cut is a $\varphi^Q\rightarrow Q \varphi$ tree form factor, the r.h.s. represents a $(L-n+1)$-loop $2n$-point scattering amplitude.}
\label{fig:phi4cut1}
\end{figure}

In each Feynman diagram contributing to the $\varphi^Q\rightarrow Q \varphi$ form factor, there is a single vertex which contains $\varphi^Q$, and we will only cut all the internal legs which are directly incident to this vertex\footnote{One may consider other types of cuts which affect legs not originating from $\varphi^Q$, but the cuts we consider are sufficient to fix the complete integrand.} (in other words, all internal legs originating form $\varphi^Q$).  After the cut, the form factor is split into a $\varphi^Q\rightarrow Q \varphi$ tree form factor, and one or more scalar amplitudes.
For example, Figure \ref{fig:phi4cut2} shows a unitarity cut of the $\varphi^5\rightarrow5\varphi$ form factor. The building blocks of the cut are a $\varphi^5\rightarrow5\varphi$ tree form factor (the blob in the middle), and two 4-point scalar tree amplitudes (the other two blobs). Figure \ref{fig:phi4cut1} shows a unitarity cut which contains two building blocks: a $\varphi^Q\rightarrow Q \varphi$ tree form factor, and a $2n$-point scalar loop amplitude.

If a unitarity cut contains two or more scalar amplitudes (besides the tree form factor), we will call it a \textbf{reducible cut}. If the cut contains a single scalar amplitudes, we will call it an \textbf{irreducible cut}. The cut in Figure \ref{fig:phi4cut2} is a reducible cut, while the cut in Figure \ref{fig:phi4cut1} is an irreducible cut.
The integrand of a reducible cut can be split into multiple pieces. For example, Figure \ref{fig:phi4cut2} corresponds to the following cut integrand\footnote{As discussed in Section \ref{subsection:O(N)}, the tree form factor equals 1. Therefore the cut integrand does not depend on the number of external legs which are originating from $\varphi^Q$.}:
\begin{equation}
I=I_1I_2,\ I_1=\frac{g}{l_1^2(l+p_{12})^2},\ I_2=\frac{g}{l_2^2(l_2+p_{34})^2}\ .
\end{equation}
Both $I_1$ and $I_2$ are identical to a cut integrand of a one-loop $\varphi^2\rightarrow 2\varphi$ form factor. In general, each reducible cut can be regarded as the product of several lower loop irreducible cuts.


It is not hard to see that:
\begin{enumerate}
\item Each Feynman diagram contains at least two internal legs which are originating from $\varphi^Q$, so it can always be captured\footnote{A Feynman diagram is captured by a cut, if all cut propagators are also propagators of this Feynman diagram. On the contrary, if a cut propagator is not the propagator of a Feynman diagram, then this Feynman diagram gives zero under the cut, which means it is not captured by the cut.} by a  reducible or an irreducible cut. Therefore these two types of cuts will suffice to fix the complete integrand.

\item Different cuts have no intersections. If a Feynman diagram is captured by one cut, then it cannot be captured by any other cuts.
\end{enumerate}
This means the loop integrand $\mathcal{F}_Q^{(L)}$ can be obtained by summing over contributions of all reducible and irreducible cuts.
More precisely, let $\{\mathcal{C}_1,\cdots \mathcal{C}_{N_c}\}$ be the complete set of reducible and irreducible cuts, and $\{l_1,\cdots, l_{n_i}\}$ be the momenta of cut legs in $\mathcal{C}_i$. The cut integrand will denoted by $\mathcal{F}_Q^{(L)}\Bigr|_{\mathcal{C}_i}$, and we define the \textbf{restored cut integrand} as
\begin{equation}
\mathcal{F}_Q^{(L)}\Bigr|_{\mathcal{C}_i}^{\text{restored}}
=\frac{\mathcal{F}_Q^{(L)}\Bigr|_{\mathcal{C}_i}}{l_1^2\cdots l_{n_i}^2}\ .
\end{equation}
The restored cut integrand is actually the contribution of all Feynman diagrams captured by the cut. Therefore
\begin{equation}
\mathcal{F}_Q^{(L)}=
\sum_{i=1}^{N_c}\mathcal{F}_Q^{(L)}\Bigr|_{\mathcal{C}_i}^{\text{restored}}\ .
\end{equation}

Let us examine the irreducible cuts in Figure \ref{fig:phi4cut1}.
The l.h.s. of the cut is $\mathcal{F}_Q^{(0)}=1$, so the cut integrand does not depend on $Q$ or $p_{n+1},\cdots, p_Q$. Therefore the corresponding restored cut integrand can be denoted by $\mathtt{F}_n^{(L)}(p_1,\cdots, p_n)$. Similar cut channels can be obtained by permuting all $Q$ external legs, and in all $C_Q^n$ different $\mathtt{F}_n^{(L)}(p_{i_1},\cdots, p_{i_n})$ are generated. However, the UV counterterm is independent of the external momenta, because the integral has divergence degree 0 (see Appendix \ref{appendixB} for details). So from now on we will not discriminate $\mathtt{F}_n^{(L)}$ with different external momenta, and their total contribution is $C_Q^n \mathtt{F}_n^{(L)}$.

The value of $n$ in Figure \ref{fig:phi4cut1} satisfies $2\le n\le L+1$, so the contribution of all irreducible cuts is given by
\begin{equation}
\mathcal{F}_Q^{(L),\text{irreducible}}= \sum_{n=2}^{L+1}C_Q^n \mathtt{F}_n^{(L)}\ .
\end{equation}

The contribution of reducible cuts can be constructed by the product of several $\mathtt{F}_{n_i}^{(L_i)}$ divided by the corresponding symmetry factor. 
The complete form factor can be written as
\begin{equation}\label{form-factor-exp}
\mathcal{F}_Q^{(L)}= \sum_{k=1}^{L}\sum_{\text{allowed }(n_i, L_i)}\frac{Q!}{(Q-\sum_{i=1}^kn_i)!}
\frac{1}{S^{L_1\cdots L_k}_{n_1\cdots n_k}}\prod_{i=1}^k\frac{1}{n_i!}\mathtt{F}_{n_i}^{(L_i)}\ ,
\end{equation}
in which the allowed $(n_i, L_i)$ should satisfy the following condition:
\begin{equation}
n_i\ge 2,\ \sum_{i=1}^k n_i\le L+k,\ \sum_{i=1}^k L_i= L,\ 
\end{equation}
and $\{(n_i,L_i)|i=1,\cdots,k\}$ should be an ordered set, which means $n_1\le n_2\le \cdots \le n_k$, and if $n_i=n_{i+1}$, then $L_i\le L_{i+1}$. 
$S^{L_1\cdots L_k}_{n_1\cdots n_k}$ is a symmetry factor when there are duplicated $(n_i,L_i)$ pairs. For example, $S^{31112}_{22335}=2!$ because there are two $(n_i,L_i)=(3,1)$ pairs. 

As an example of \eqref{form-factor-exp}, the $L=3$ expression is
\begin{equation}
\begin{aligned}
&\mathcal{F}_Q^{(3)}=C_Q^2\mathtt{F}_2^{(3)}+C_Q^3\mathtt{F}_3^{(3)}
+C_Q^4\mathtt{F}_4^{(3)}+6C_Q^4\mathtt{F}_2^{(1)}\mathtt{F}_2^{(2)}
+10C_Q^5\mathtt{F}_2^{(1)}\mathtt{F}_3^{(2)}
+15C_Q^6\Bigl[\mathtt{F}_2^{(1)}\Bigr]^3\ .\\
\end{aligned}\label{FQ3}
\end{equation}

\begin{figure}[htb]
\centering
\includegraphics[scale=0.45]{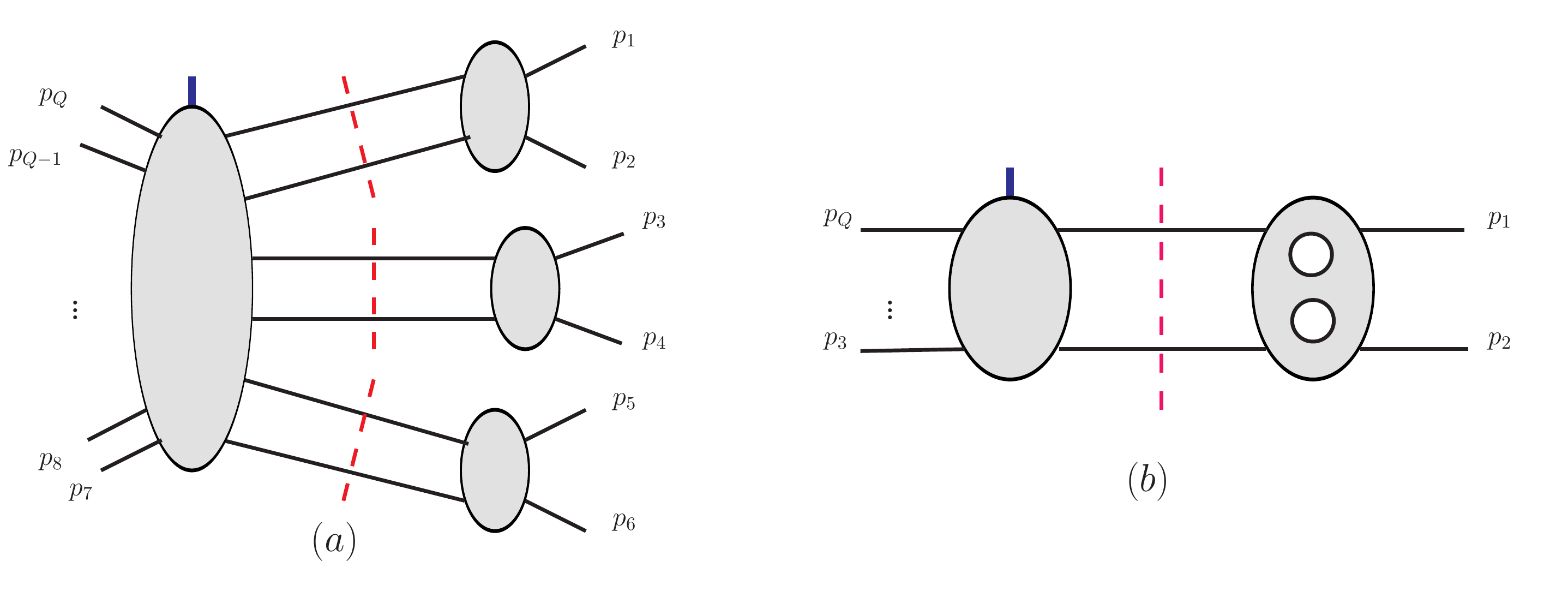}
\caption{The cuts corresponding to the last term and the first term in \eqref{FQ3}.}
\label{fig:phi4cut3}
\end{figure}

The last term in \eqref{FQ3} corresponds to the reducible cut in Figure \ref{fig:phi4cut3} (a). It has $(n_i,L_i)=(2,1)$ for $i=1,2,3$, so the factor $S_{222}^{111}=3!$. The overall factor for this cut is
\begin{equation}
\frac{Q!}{(Q-6)!}\frac{1}{S^{111}_{222}(2!)^3}
=\frac{Q!}{(Q-6)!}\frac{1}{3!(2!)^3}=15C_Q^6\ .
\end{equation}
The first term in \eqref{FQ3} corresponds to the irreducible cut in Figure \ref{fig:phi4cut3} (b), and the corresponding factor is $\frac{Q!}{(Q-2)!}\frac{1}{2!}=C_Q^2$.

From \eqref{form-factor-exp} and \eqref{FQ3} we observe that for a given $L$, $\mathcal{F}_Q^{(L)}$ can be constructed by the following building blocks: \begin{equation}\label{F-building-block}
\mathcal{F}_Q^{(L)}:\ \{\mathtt{F}_n^{(L_1)}|n=2,\cdots, L+1; L_1=n-1,\cdots, L\}\ .
\end{equation}
Therefore the complexity of $\mathcal{F}_Q^{(L)}$ only depends on $L$, while $Q$ only appears in the overall factors.

\subsection{Integrand of $2n$-point scattering amplitudes}
\label{subsection:2n-amplitude}
The r.h.s. of the unitarity cut in Figure \ref{fig:phi4cut1} is a $2n$-point $(L-n+1)$-loop scattering amplitude $\mathcal{A}_{2n}^{(L-n+1)}$. 
The list of cut integrands and scattering amplitudes needed to computed $\mathcal{F}_Q^{(L)}$ to 5-loop are given in Table \ref{cutlist}.
\begin{table}
\begin{tabular}{|c |c|c|c|c|c|}
\hline
$\mathtt{F}_n^{(L)}$ & $\mathtt{F}_2^{(1)}$& $\mathtt{F}_2^{(2)}$, $\mathtt{F}_3^{(2)}$&  $\mathtt{F}_2^{(3)}$, $\mathtt{F}_3^{(3)}$, $\mathtt{F}_4^{(3)}$&  $\mathtt{F}_2^{(4)}$, $\mathtt{F}_3^{(4)}$, $\mathtt{F}_4^{(4)}$, $\mathtt{F}_5^{(4)}$&  $\mathtt{F}_2^{(5)}$, $\mathtt{F}_3^{(5)}$, $\mathtt{F}_4^{(5)}$, $\mathtt{F}_5^{(5)}$, $\mathtt{F}_6^{(5)}$  \\
\hline
$\mathcal{A}_n^{(L)}$ &$\mathcal{A}_4^{(0)}$&$\mathcal{A}_4^{(1)}$, $\mathcal{A}_6^{(0)}$&$\mathcal{A}_4^{(2)}$, $\mathcal{A}_6^{(1)}$, $\mathcal{A}_8^{(0)}$&$\mathcal{A}_4^{(3)}$, $\mathcal{A}_6^{(2)}$, $\mathcal{A}_8^{(1)}$, $\mathcal{A}_{10}^{(0)}$&$\mathcal{A}_4^{(4)}$, $\mathcal{A}_6^{(3)}$, $\mathcal{A}_8^{(2)}$, $\mathcal{A}_{10}^{(1)}$, $\mathcal{A}_{12}^{(0)}$\\
\hline
\end{tabular}
\caption{Cut integrands and scattering amplitudes required to compute $\mathcal{F}_Q^{(L)}$ to 5-loop.}
\label{cutlist}
\end{table}

When $n$ or $L$ are large, it can be very difficult to generate the complete set of Feynman diagrams contributing to $\mathcal{A}_{2n}^{(L-n+1)}$. 
Fortunately, as discussed earlier, all Feynman diagrams with the same topology have the same contribution to the UV counterterm. 
Therefore we only need to generate all different topologies, and determine the necessary combinatorial factors from the permutation of external legs.

\begin{figure}[htb]
\centering
\includegraphics[scale=0.6]{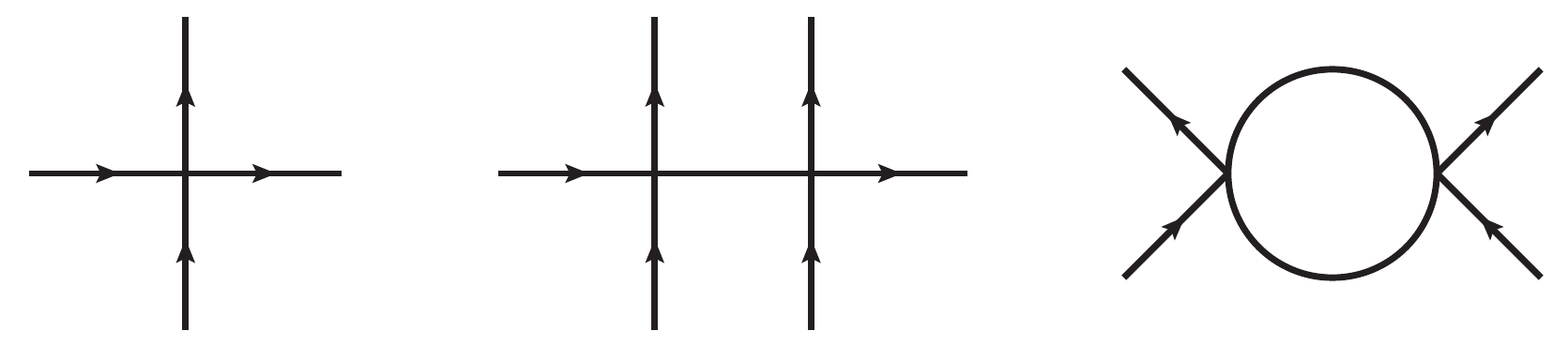}
\caption{Some $\mathcal{A}_{2n}^{(L)}$ topologies.}
\label{fig:varphi-topo}
\end{figure}

Some $\mathcal{A}_{2n}^{(L)}$ topologies are shown in Figure \ref{fig:varphi-topo}, in which we have added outgoing(incoming) arrows to $\varphi$($\bar{\varphi}$) external legs. These topologies can be constructed by finding all inequivalent ways of adding arrows to the external legs in $\phi^4$ topologies.

\begin{figure}[htb]
\centering
\includegraphics[scale=0.45]{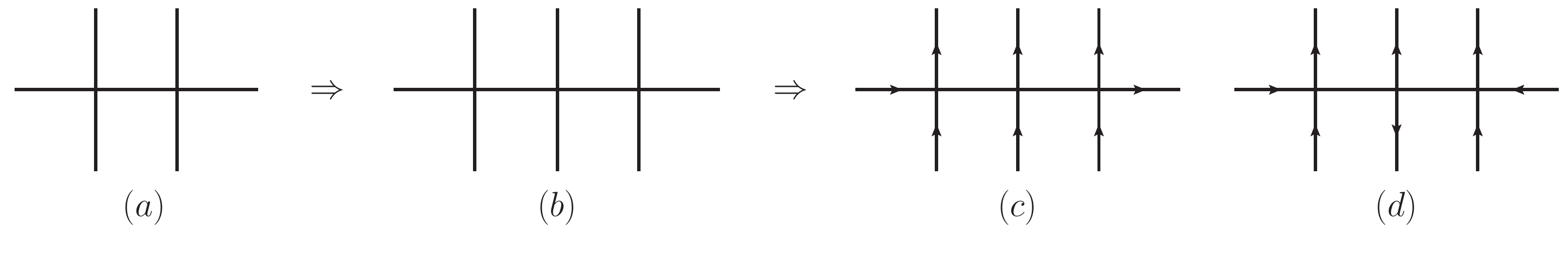}
\caption{(a): the 6-point $\phi^4$ topology. 
(b): the 8-point $\phi^4$ topology.
(c) and (d): the topologies contributing to $\mathcal{A}_{8}^{(0)}$. }
\label{fig:tree-topo-gen}
\end{figure}

The $\phi^4$ topologies can be constructed recursively by adding vertices to or gluing external legs in lower point or lower loop topologies.
The $(2n+2)$-point tree topologies can be generated by finding all inequivalent ways of adding a $\phi^4$ vertex to $2n$-point topologies. An example is given in Figure \ref{fig:tree-topo-gen}.

\begin{figure}[htb]
\centering
\includegraphics[scale=0.45]{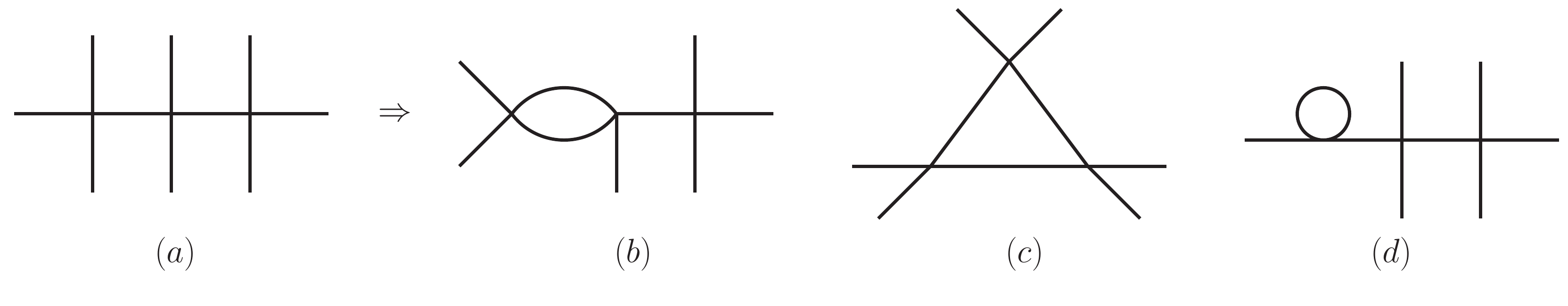}
\caption{From 8-point tree topology to 6-point 1-loop topologies. (d) should be removed because it contains a snail subgraph, therefore it has no contribution to the UV counterterm. }
\label{fig:loop-topo-gen}
\end{figure}

The $L$-loop $2n$-point $\phi^4$ topologies can be generated by gluing two external legs together in the $(L-1)$-loop $(2n+2)$-point $\phi^4$ topologies, and removing graphs containing snail subgraphs. An example is given in Figure \ref{fig:loop-topo-gen}.

The integrand of $\mathcal{A}_{2n}^{(L-n+1)}$ can be constructed from the topologies, together with some combinatorial factors.
Let $T$ be a topology of $\mathcal{A}_{2n}^{(L-n+1)}$, and let $\{G_i(T)|i=1,\cdots,N_T\}$ be the Feynman diagrams corresponds to $T$. 
$N_T$ depends on the symmetry factor $S_{\varphi}$($S_{\bar{\varphi}}$) of the diagram under permutation of $n$ $\varphi$($\bar{\varphi}$) external legs:
\begin{equation}
N_T=\frac{(n!)^2}{S_{\varphi}S_{\bar{\varphi}}}\ .
\end{equation}
The integrand corresponding to $G_i(T)$ will be denoted by the same $I(T)$, then the total integrand can be written as
\begin{equation}\label{A2n}
\mathcal{A}_{2n}^{(L-n+1)}=\sum_T \frac{(n!)^2}{S^{int}_TS_{\varphi}S_{\bar{\varphi}}}I(T)\ ,
\end{equation}
in which $S^{int}_T$ is the symmetry factor of the diagram under permutation of internal legs.

We can also define $S_T\equiv S^{int}_TS_{\varphi}S_{\bar{\varphi}}$, which is the symmetry factor of $T$ under the permutations of internal and external legs all together. Then \eqref{A2n} becomes
\begin{equation}
\mathcal{A}_{2n}^{(L-n+1)}=\sum_T \frac{(n!)^2}{S_T}I(T)\ .
\end{equation}
The restored cut integrand of the form factor is
\begin{equation}
\mathtt{F}_{n}^{(L)}=\frac{1}{l_1^2\cdots l_n^2}\sum_T \frac{n!}{S_T}I(T)\ ,
\end{equation}
in which we have added the cut propagators and removed a $n!$ factor to compensate the symmetry factor of $n$ cut legs.

We also generated some loop form factors directly using QGRAF \cite{Nogueira:1991ex}, including all form factors with $Q\le 6, L\le 3$, and $\varphi^2\rightarrow 2\varphi$, $\varphi^3\rightarrow 3\varphi$ to 5 loop. All the results are consistent with the unitarity cut method. 

\section{The anomalous dimensions}
\label{section:anomalous}
The bare operator is defined as $\varphi^Q_{bare}\equiv Z_{\varphi^Q}\varphi^Q$, and the renormalization Z-factor can be determined by the one-particle-irreducible (1PI) part of the form factor, $\mathcal{F}_{Q}^{(L),1PI}$.
Unitarity cut method produces the complete scattering amplitudes (form factors) including the contributions of both 1PI and one-particle-reducible(1PR) diagrams, but the contribution of 1PR diagrams can be easily removed in a pure scalar theory\footnote{For $\varphi^Q\rightarrow Q\varphi$ form factor, 1PR diagrams with $\frac{1}{(p_1+p_2+\cdots+p_n)^2}$ type poles are forbidden by charge conservation unless $n=1$. So we only need to remove the diagrams with $\frac{1}{p_i^2}$ type propagators}.

\begin{figure}[htb]
\centering
\includegraphics[scale=0.7]{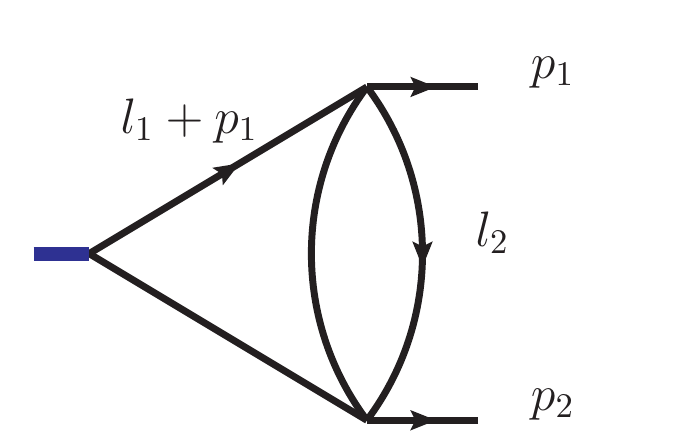}
\caption{An integral which contributes to $\mathtt{F}_{2}^{(2)}$. The arrows in the diagram represent the directions of the momenta.}
\label{fig:uv-example-2}
\end{figure}

We evaluate the UV divergences of the form factors using the UV decomposition method.  The UV divergences of multiloop integrals can be decomposed into the local divergence and lower loop sub divergences, and the anomalous dimensions are determined solely by the local divergences. 
Here we only demonstrate the method by considering a two loop integral which corresponds to the Feynman diagram in Figure \ref{fig:uv-example-2}. More details can be found in \cite{Jin:2022zcj}.

The local divergence of a generic integral can be reduced to that of vacuum integrals by an asymptotic expansion around large loop momenta $l_i$.  Since this integral has divergence degree 0, we can simply set $p_i\rightarrow0$ without changing the local divergence:
\begin{equation}
\begin{aligned}
&\mathbf{L}\frac{1}{(l_1+p_1)^2(l_1-p_2)^2(l_1-l_2)^2l_2^2}
=\mathbf{L}\frac{1}{(l_1^2)^2(l_1-l_2)^2l_2^2}\ ,\\
\end{aligned}
\end{equation}
in which $\mathbf{L}$ is the local divergence operator, and we have ignored the integration symbol. Then the infrared divergence of the vacuum integral is regulated by adding mass to a propagator:
\begin{equation}
\begin{aligned}
&\mathbf{L}\frac{1}{(l_1^2)^2(l_1-l_2)^2l_2^2}
=\mathbf{L}\frac{1}{(l_1^2+m^2)^2(l_1-l_2)^2l_2^2}\ .\\
\end{aligned}
\end{equation}
The local divergence of the massive vacuum integral can be obtained by subtracting sub UV divergences from the total UV divergence of the integral\footnote{The local divergence is different from \cite{Jin:2022zcj} because the different conventions of $\epsilon$. In this work $D=4-\epsilon$, but in \cite{Jin:2022zcj} $D=4-2\epsilon$. The $e^{\epsilon\gamma}$ factor in \eqref{local-div-3} is from the convention of $\overline{\text{MS}}$ scheme.}:
\begin{equation}
\begin{aligned}
&\mathbf{L}\frac{1}{(l_1^2+m^2)^2(l_1-l_2)^2l_2^2}
= \frac{1}{(l_1^2+m^2)^2(l_1-l_2)^2l_2^2}
-\frac{1}{(l_1^2+m^2)^2}\mathbf{L}\frac{1}{(l_1-l_2)^2l_2^2}\\
=&\frac{m^{2d-8}e^{\epsilon\gamma}\Gamma(4-d)\Gamma(2-\frac{d}{2})\Gamma(-1+\frac{d}{2})^{2}}{\Gamma(\frac{d}{2})}
-\frac{e^{\frac{1}{2}\epsilon\gamma}}{\epsilon}m^{d-4}\Gamma(2-\frac{d}{2})
=-\frac{2}{\epsilon ^{2}}+\frac{1}{\epsilon }\ .\\
\end{aligned}\label{local-div-3}
\end{equation}

The Z-factor is given by the local divergence of the form factor:
\begin{equation}\label{zq1}
Z_{\varphi^Q}=1-\sum_{L=1}^{\infty}\mathbf{L}\mathcal{F}_{Q}^{(L),1PI}\ .
\end{equation}
The UV decomposition method turns out to be quite efficient even at 5-loop, and we were able to evaluate the local divergences of hundreds of 5-loop integrals in a few hours on a laptop computer.

The anomalous dimensions $\gamma_{\varphi^Q}$ can be derived from the combination $\mathcal{Z}_Q\equiv Z_{\phi}^{-\frac{Q}{2}}Z_{\varphi^Q}$:
\begin{equation}
\begin{aligned}
\gamma_{\varphi^Q}
=&-\frac{\partial \ln \mathcal{Z}_Q}{\partial\ln \mu}
=\Bigl[\epsilon g-\beta(g)\Bigr]\mathcal{Z}_Q^{-1}\frac{\partial \mathcal{Z}_Q}{\partial g}
=Q\sum_{L=1}^{\infty}g^L \gamma_L\ .\\
\end{aligned}\label{anomalous-dimension-define}
\end{equation}
The wavefunction renormalization factor $Z_{\phi}$ and the beta function $\beta(g)$ can be found in \cite{Kleinert:1991rg}, and we have also reproduced them using the UV decomposition method.

The 1PI form factors vanishes when $Q=1$, but \eqref{anomalous-dimension-define} still holds:
\begin{equation}
\gamma_{\varphi^1}
=-\frac{\partial \ln Z_{\phi}^{-\frac{1}{2}}}{\partial\ln \mu}
=\frac{1}{2}\frac{\partial \ln Z_{\phi}}{\partial\ln \mu}\ .
\end{equation}

The $\gamma_L$ in \eqref{anomalous-dimension-define} are given by
\begin{equation}
\begin{aligned}
\gamma_1=&\frac{Q-1}{3},\ \\
\gamma_2=&\frac{2Q^{2}}{9}+\frac{(N-6)Q}{18}+\frac{2-3N}{36},\ \\
\gamma_3=&\frac{8Q^{3}}{27}
+Q^{2}\Bigl[\frac{N-36}{27}+\frac{2(N+14)\zeta_{3}}{27}\Bigr]
+Q\Bigl[\frac{646+23N-2N^{2}}{216}
-\frac{2(N+14)\zeta_{3}}{9}\Bigr]\\
&+\frac{3N^{2}-72N-860}{432}+\frac{4(N+14)\zeta_{3}}{27},\ \\
\end{aligned}\label{anomalous-dimension-result-1}
\end{equation}
\begin{equation}
\begin{aligned}
\gamma_4=&\frac{14Q^{4}}{27}
+Q^{3}\Bigl[\frac{4(N-73)}{81}+\frac{2(6N+65)\zeta_{3}}{81}+\frac{5(N+30)\zeta_{5}}{81}\Bigr]\\
&+Q^{2}\Bigl[\frac{(1634-15N-2N^{2})}{162}
+\frac{(3N^{2}-14N-360)\zeta_{3}}{81}
-\frac{(N+8)(N+14)\pi^{4}}{4860}\\
&\ \ \ \ \ \ \ \ \ +\frac{70(N-2)\zeta_{5}}{81}\Bigr]\\
&+Q\Bigl[\frac{3N^{3}+137N^{2}+2448N-16748}{1944}
-\frac{(N^{3}+36N^{2}+168N-816)\zeta_{3}}{324}\\
&\ \ \ \ \ \ \ \ +\frac{(3N^{2}+76N+380)\pi^{4}}{4860}
-\frac{35(7N+18)\zeta_{5}}{81}\Bigr]\\
&-\frac{9N^{3}+1144N^{2}+20952N-22120}{15552}
+\frac{(N^{3}+24N^{2}+176N+104)\zeta_{3}}{324}\\
&-\frac{(N^{2}+27N+134)\pi^{4}}{2430}+\frac{10(17N+62)\zeta_{5}}{81}\ .\\
\end{aligned}\label{anomalous-dimension-result-2}
\end{equation}
The expression of $\gamma_5$ can be found in Appendix \ref{appendixC}.

The Wilson-Fisher fixed point is define by
\begin{equation}
\frac{\partial g}{\partial\ln \mu}\Bigr|_{g=g*}=\epsilon g^*-\beta(g^*)=0\ .
\end{equation}
At the Wilson-Fisher fixed point, the scaling dimension of  $\varphi^Q$ can be expanded into series of $\epsilon$, 
\begin{equation}
\triangle_{\varphi^Q}=Q(1-\frac{\epsilon}{2})+\gamma_{{\varphi^Q}}\Bigr|_{g=g^*}
\equiv Q+Q\sum_{l=1}^{\infty}\epsilon^l P_l(Q)\ ,
\end{equation}
in which $Q(1-\frac{\epsilon}{2})$ is the classical scaling dimension of $\varphi^Q$ in the $4-\epsilon$ dimensional space.

Our result for $\triangle_{\varphi^Q}$ is in complete agreement with \cite{Antipin:2020abu}, in which $P_1(Q)$ to $P_4(Q)$ are fixed using both semiclassical and known perturbative results at $Q=1,2,4$. The 5-loop contribution $P_5(Q)\equiv\sum_{i=0}^5P_{5}^i Q^i$ is given by
\begin{equation}
\begin{aligned}
P_{5}^5=&\frac{256}{(8+N)^{5}},\\
P_{5}^4=&\frac{3(7N^{2}-1248N-8752)}{(8+N)^{6}}+\frac{28(28+3N)\zeta_{3}}{(8+N)^{5}}
+\frac{40(22+N)\zeta_{5}}{(8+N)^{5}}+\frac{14(62+N)\zeta_{7}}{(8+N)^{5}},\\
P_{5}^3=&-\frac{6(-173104-48736N-3036N^{2}+73N^{3}+N^{4})}{(8+N)^{7}}
+\frac{7(206+361N)\zeta_{7}}{4(8+N)^{5}}\\
&+\frac{(18N^{3}-673N^{2}-15210N-60128)\zeta_{3}}{(8+N)^{6}}
-\frac{(65+6N)\pi^{4}}{60(8+N)^{4}}
-\frac{5(30+N)\pi^{6}}{756(8+N)^{4}}\\
&+\frac{(1104+158N+13N^{2})\zeta_{3}^{2}}{2(8+N)^{5}}
+\frac{5(-16736-1744N+166N^{2}+5N^{3})\zeta_{5}}{2(8+N)^{6}},\\
P_{5}^2=&\frac{-19621472-8157920N-1037502N^{2}-19068N^{3}+3419N^{4}
+\frac{291N^{5}}{4}+\frac{N^{6}}{2}}{(8+N)^{8}}\\
&-\frac{(-3372512-1321456N-134112N^{2}+1116N^{3}+362N^{4}+N^{5})\zeta_{3}}{2(8+N)^{7}}\\
&-\frac{(-7464-1720N-44N^{2}+3N^{3})\pi^{4}}{120(8+N)^{5}}
+\frac{49(-76+4N+9N^{2})\zeta_{7}}{4(8+N)^{5}}\\
&-\frac{(37472+12328N+1258N^{2}+53N^{3})\zeta_{3}^{2}}{(8+N)^{6}}
-\frac{5(-2+N)\pi^{6}}{54(8+N)^{4}}\\
&+\frac{(86496-26060N-4324N^{2}+301N^{3}+2N^{4})\zeta_{5}}{2(8+N)^{6}},\\
\end{aligned}\label{p55}
\end{equation}

\begin{equation}
\begin{aligned}
P_{5}^1=&-\frac{1}{64(8+N)^{9}}\Bigl(-11161819136-6069470208N-1140742656N^{2}\\
&-61478976N^{3}+5999960N^{4}+824504N^{5}+25772N^{6}+463N^{7}+4N^{8}\Bigr)\\
&+\frac{(-514112-181184N-12224N^{2}+904N^{3}+52N^{4}+N^{5})\pi^{4}}{480(8+N)^{6}}\\
&+\frac{5(2496+958N+65N^{2})\pi^{6}}{756(8+N)^{5}}\\
&-\frac{\zeta_{3}}{8(8+N)^{8}}\Bigl(159301888+82561536N+12675296N^{2}+111872N^{3}\\
&-94960N^{4}-3608N^{5}-48N^{6}+N^{7}\Bigr)\\
&+\frac{(1491072+729440N+116880N^{2}+7554N^{3}+247N^{4})\zeta_{3}^{2}}{2(8+N)^{7}}\\
&-\frac{(-1323968-1009544N-77280N^{2}+22856N^{3}+2021N^{4}+6N^{5})\zeta_{5}}{2(8+N)^{7}}\\
&-\frac{7(190840+76762N+7771N^{2}+189N^{3})\zeta_{7}}{4(8+N)^{6}}\ ,\\
\end{aligned}\label{p51}
\end{equation}

\begin{equation}
\begin{aligned}
P_{5}^0=&\frac{1}{256(8+N)^{9}}\Bigl(-18215837696-9842300928N-1755406336N^{2}\\
&-53369344N^{3}+18477984N^{4}+1992280N^{5}+47688N^{6}+648N^{7}
+3N^{8}\Bigr)\\
&-\frac{(-300096-100480N-3496N^{2}+1056N^{3}+40N^{4}+N^{5})\pi^{4}}{480(8+N)^{6}}\\
&+\frac{\zeta_{3}}{16(8+N)^{8}}\Bigl(155171328+78629632N+10305472N^{2}-485824N^{3}\\
&-141784N^{4}-3208N^{5}-68N^{6}+3N^{7}\Bigr)
-\frac{5(1240+418N+25N^{2})\pi^{6}}{378(8+N)^{5}}\\
&-\frac{(481088+242512N+38280N^{2}+2278N^{3}+77N^{4})\zeta_{3}^{2}}{(8+N)^{7}}\\
&+\frac{(-1340096-657800N+13132N^{2}+26662N^{3}+1759N^{4}+4N^{5})\zeta_{5}}{2(8+N)^{7}}\\
&+\frac{21(31580+12236N+1145N^{2}+21N^{3})\zeta_{7}}{2(8+N)^{6}}\ .\\
\end{aligned}
\end{equation}
The $P_{55}$ and $P_{54}$ in \eqref{p55} are also in agreement with the semiclassical computation in \cite{Antipin:2020abu}. The anomalous dimension and scaling dimension at the Wilson-Fisher fixed point are also available in ancillary files to the arXiv submission of this article.

In \cite{Badel:2019oxl}, the scaling dimension in the $\phi^4$ theory with $U(1)$ symmetry was computed to 5-loop by combining semiclassical and perturbative results. 
This scaling dimension can also be obtained by setting $N=2$ in our result. We find that $P_1$ to $P_3$ in \cite{Badel:2019oxl} are consistent with our result.
$P_4$ is different, which stems from a typo in \cite{Calabrese:2002bm}, as was pointed out in \cite{Antipin:2020abu}. 
$P_5$ was given in numerical form in \cite{Badel:2019oxl}, and it is in agreement with our result.

Recently, the complete $Q$ dependence of $P_5(Q)$ was also obtained in \cite{Henriksson:2022rnm} by combining the results of semi-classical and perturbative computations\footnote{The $Q=3$ results was extracted from the general results in \cite{Bednyakov:2021ojn}.}, and it was in complete agreement with our result.

\section{Discussion}
\label{section:discussion}

In this work we have computed the anomalous dimensions of the $\phi^Q$-type operators to 5-loop. It is worth extending the computation to 6-loop, which gives further check to the scaling dimensions computed using the semiclassical method, and also allows us to test the efficiency of the UV decomposition method beyond 5-loop.  
It should not be hard to generate the 6-loop form factor integrand using the unitarity cut method, and the analytic expressions of propagator type master integrals needed for the computation of 6-loop UV divergence can be found in \cite{Georgoudis:2021onj}. 

The large-$Q$ scaling dimensions in a $\phi^4$ theory with $U(N)\times U(N)$ symmetry were computed using semiclassical methods in \cite{Antipin:2020rdw}, and the results were verified using direct Feynman diagram computation to 4-loop in \cite{Jack:2021lja}. With the help of unitarity cut and UV decomposition method, one should be able to obtain the full $Q$-dependence of scaling dimensions to 5-loop.

The same methods may also be applied to other operators, like the $\partial^{2m}\phi^n$-type operators considered in \cite{Cao:2021cdt}, and the operators in more complicated $O(N)$ representations considered in \cite{Kehrein:1995ia}.

\acknowledgments

We would like to thank Bo Feng, Johan Henriksson, Hui Luo, Rui Yu, Gang Yang for helpful discussions.



\appendix

\section{The factorization of $O(N)$ tensor structure}
\label{appendixA}
 In each loop Feynman diagram contributing to the $\phi^Q\rightarrow Q\phi$ form factor, the tree form factor can be factored out, and the integrand is free of $O(N)$ indices. To understand this feature, let us rewrite the Lagrangian with the help of an auxiliary field $\sigma$,
\begin{equation}\label{cubicLagrangian}
\mathcal{L}\rightarrow \frac{1}{2}(\partial_{\mu}\phi)^2-\frac{1}{2}\sigma^2+\sqrt{\frac{g}{12}}\sigma\phi^2\ .
\end{equation}
The original Lagrangian is recovered when the auxiliary field is integrated out.

\begin{figure}[htb]
\centering
\includegraphics[scale=0.5]{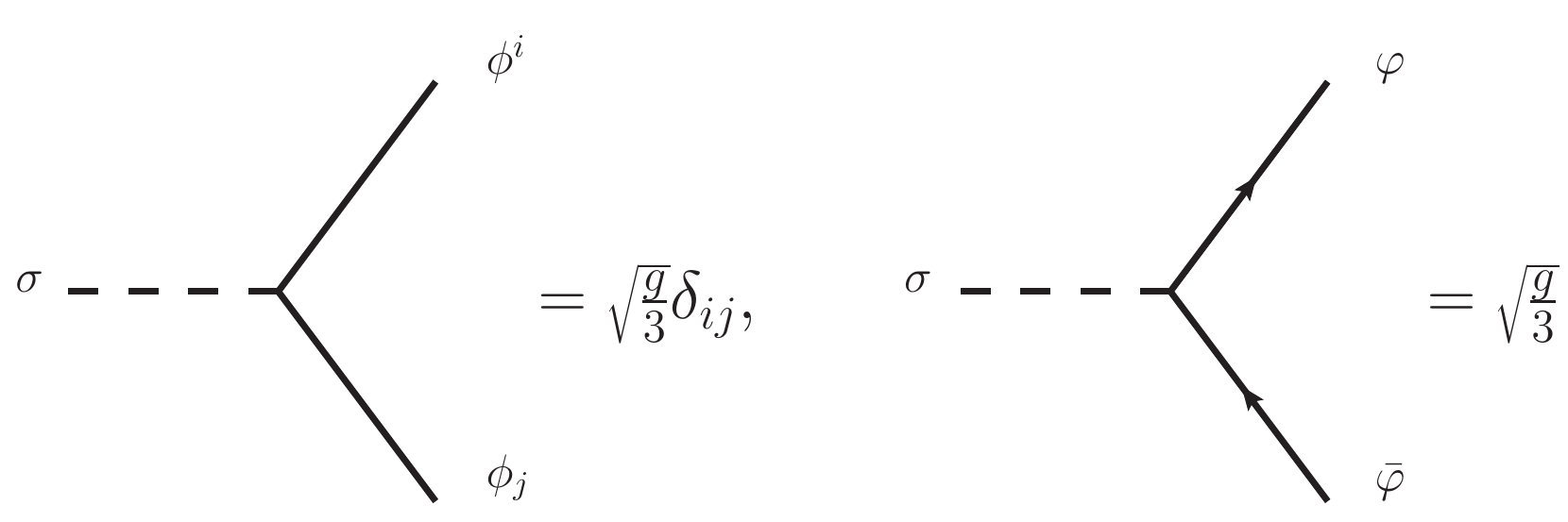}
\caption{The interaction vertices of the Lagrangian given by \eqref{cubicLagrangian}.}
\label{fig:sigmaphiphivertex}
\end{figure}

In the new Lagrangian, $\sigma$ is a $O(N)$ singlet, and the interaction vertices are shown in Figure \ref{fig:sigmaphiphivertex}. In the Figure, dashed lines represents $\sigma$, solid lines without arrows represents $\phi^i$, and solid lines with arrows represents $\varphi\equiv q\cdot\phi$ and $\bar{\varphi}$.

\begin{figure}[htb]
\centering
\includegraphics[scale=0.7]{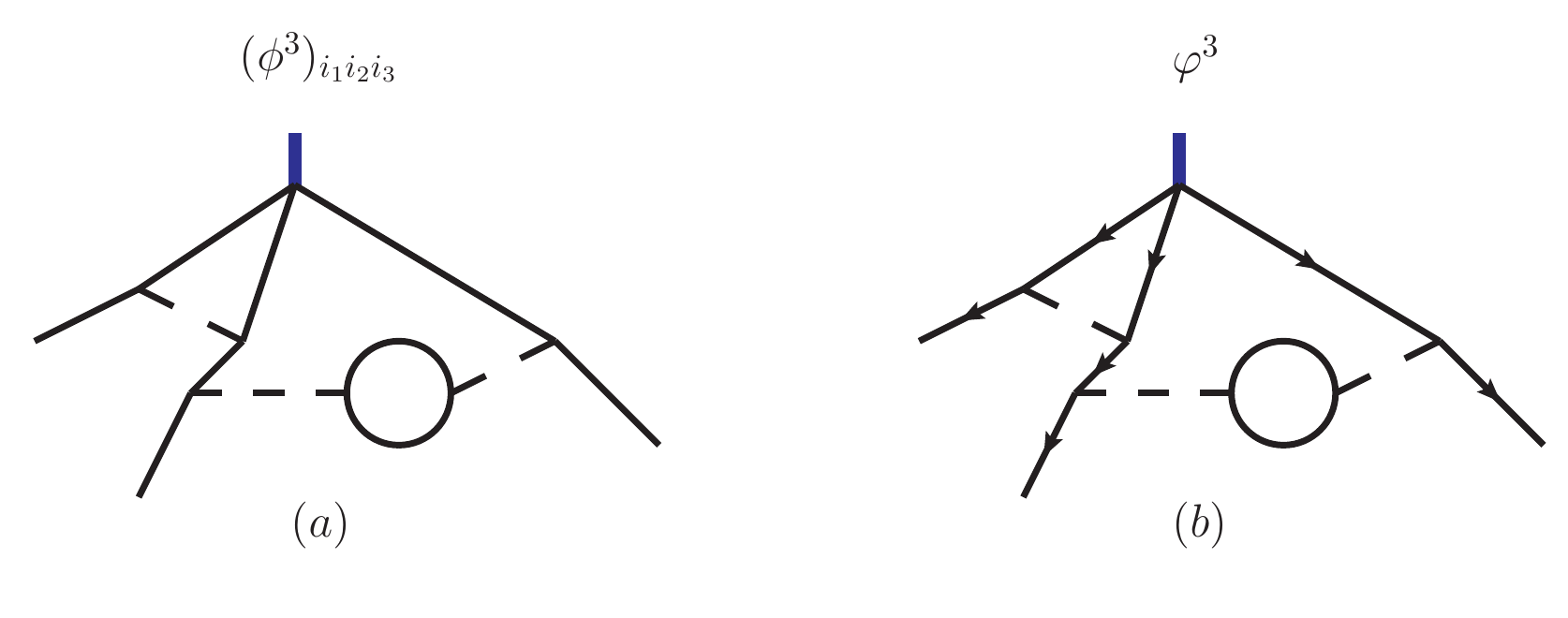}
\caption{Feynman diagram contributing to (a) $\phi^Q\rightarrow Q\phi$ and (b) $\varphi^Q\rightarrow Q\varphi$.}
\label{fig:phi3graph}
\end{figure}

An example of Feynman diagrams contributing to $\phi^Q\rightarrow Q\phi$ is shown in Figure \ref{fig:phi3graph}(a). In these diagrams, $\phi$-fields either form a closed $\phi$-loop, or a $\phi$-line connecting two external legs. The $\phi$-lines attached to the operator will not  form $\phi$-loops, because the operator is traceless.

The $O(N)$ tensor structure of the diagram can be easily read from the diagram. Each $\phi$-loop contributes a $N$ factor, and each $\phi$-line contributes a $\delta_{ij}$ factor. So it is clear that the tensor structure is proportional to the Feynman rule of the operator.

An example of Feynman diagrams contributing to $\varphi^Q\rightarrow Q\varphi$ is shown in Figure \ref{fig:phi3graph}(b). These diagrams only contain $\varphi$-lines and $\phi$-loops.

\section{The superficial degree of divergence of $\phi^Q\rightarrow Q\phi$ form factor}
\label{appendixB}

An important property of $\mathcal{F}_Q^{(L)}$ is that the superficial degree of UV divergence is 0. To see this, let $V$ and $E$ be the number of vertices and propagators of a Feynman diagram contributing to $\mathcal{F}_Q^{(L)}$. The diagram has $(V-1)$ 4-vertices, a single $(Q+1)$-vertex, and $Q+1$ external legs, so the number of propagators of the diagram is
\begin{equation}\label{degree-1}
E=\frac{4(V-1)+(Q+1)-(Q+1)}{2}=2(V-1)\ .
\end{equation}
A connected Feynman diagram also satisfies
\begin{equation}\label{degree-2}
V-E+L=1\ .
\end{equation}
Combining \eqref{degree-1} and \eqref{degree-2}, we derive $E=2L$. 
$E$ equals the number of loop propagators in the integral, therefore the superficial degree of divergence in 4-D is given by
\begin{equation}
\omega=4L-2E=0\ .
\end{equation}

The renormalization Z-factors only depend on the local UV divergences of integrals, which have polynomial dependence on external momenta (see e.g. \cite{Jin:2022zcj}). When $\omega=0$, the local UV divergence does not depend on the external momenta:
\begin{equation}
\mathbf{L}\mathcal{F}_Q^{(L)}(p_1,\cdots, p_Q)
=\mathbf{L}\mathcal{F}_Q^{(L)}(0,\cdots, 0)\ .
\end{equation}

\section{The 5-loop corrections to the anomalous dimension}
\label{appendixC}
The expression of $\gamma_5$ in \eqref{anomalous-dimension-define} is given by: 
\begin{equation}
\begin{aligned}
\gamma_5\equiv &\sum_{i=0}^5\gamma_{5i} Q^i\ ,\ \ \ \ 
\gamma_{55}=\frac{256}{243},\\
\gamma_{54}=&\frac{7N-800}{81}+\frac{28(3N+28)\zeta_{3}}{243}+\frac{40(N+22)\zeta_{5}}{243}+\frac{14(N+62)\zeta_{7}}{243}\ ,\\
\gamma_{53}=&\frac{3074-33N-2N^{2}}{81}
+\frac{(18N^{2}-385N-3994)\zeta_{3}}{243}-\frac{(N+8)(6N+65)\pi^{4}}{14580}\\
& -\frac{5(N+8)(N+30)\pi^{6}}{183708}+\frac{(13N^{2}+158N+1104)\zeta_{3}^{2}}{486}\\
&+\frac{5(5N^{2}+198N-832)\zeta_{5}}{486}+\frac{7(361N+206)\zeta_{7}}{972}\ ,\\
\gamma_{52}=&\frac{12N^{3}+881N^{2}+10230N-407000}{5832}
-\frac{(3N^{3}+28N^{2}-216N-2056)\pi^{4}}{29160}\\
&+\frac{(41348+5776N-161N^{2}-2N^{3})\zeta_{3}}{972}
+\frac{(12372-924N+285N^{2}+2N^{3})\zeta_{5}}{486}\\
&-\frac{5(N-2)(N+8)\pi^{6}}{13122}
-\frac{(53N^{2}+474N+1912)\zeta_{3}^{2}}{243}\\
&+\frac{49(-76+4N+9N^{2})\zeta_{7}}{972}\ ,\\
\gamma_{51}=&\frac{14368320+669056N+37734N^{2}-2981N^{3}-48N^{4}}{186624}\\
&+\frac{(-205528-37536N+868N^{2}-9N^{3}-2N^{4})\zeta_{3}}{3888}\\
&+\frac{(10136+5960N+878N^{2}+44N^{3}+N^{4})\pi^{4}}{116640}\\
&+\frac{5(1380+628N+53N^{2})\pi^{6}}{183708}
-\frac{(34252+5740N+1155N^{2}+6N^{3})\zeta_{5}}{486}\\
&+\frac{(2580+1376N+211N^{2})\zeta_{3}^{2}}{486}
-\frac{7(7286+2731N+189N^{2})\zeta_{7}}{972}\ ,\\
\end{aligned}\label{anomalous-dimension-result-3}
\end{equation}
\begin{equation}
\begin{aligned}
\gamma_{50}=&\frac{-6934336-1059072N-84662N^{2}+2223N^{3}+9N^{4}}{186624}\\
&+\frac{(91864+19560N-460N^{2}+13N^{3}+3N^{4})\zeta_{3}}{3888}
-\frac{5(682+253N+19N^{2})\pi^{6}}{91854}\\
&+\frac{(-14552-6176N-758N^{2}-32N^{3}-N^{4})\pi^{4}}{116640}
+\frac{(70-293N-59N^{2})\zeta_{3}^{2}}{243}\\
&+\frac{(24280+5594N+845N^{2}+4N^{3})\zeta_{5}}{486}
+\frac{7(1186+389N+21N^{2})\zeta_{7}}{162}\ .\\
\end{aligned}
\end{equation}

\bibliographystyle{JHEP}
\bibliography{/Users/jin/Documents/tex/PSUThesis/Biblio-Database}{}

\end{document}